\documentclass[aps,prapplied,reprint,amsmath,amssymb]{revtex4-2}
\usepackage{graphicx}
\usepackage{siunitx}
\usepackage{bm}
\usepackage{booktabs}
\usepackage{tabularx}
\usepackage[caption=false]{subfig}
\usepackage{placeins}
\usepackage[colorlinks=true,linkcolor=blue,citecolor=blue,urlcolor=blue]{hyperref}

\begin{document}
\title{Temperature-Dependent Charge Transport in USD-Grown High-Purity Germanium: Interplay Between Freeze-Out and Multi-Scattering Mechanisms}
\author{Narayan Budhathoki$^{1}$}
\author{Dongming Mei$^{1}$}
\email{Dongming.Mei@usd.edu}  
\author{Abhinna Rajbanshi$^{2}$}
\author{Rongying Jin$^{2}$}
\affiliation{$^{1}$Department of Physics, University of South Dakota, Vermillion, SD 57069, USA}
\affiliation{$^{2}$Department of Physics and Astronomy, University of South Carolina, Columbia, SC 29208, USA}
\date{\today}

\begin{abstract}
We report temperature-dependent charge transport measurements in p-type high-resistivity germanium crystals grown at the University of South Dakota. Hall-effect and four-probe resistivity measurements were performed on five planar samples over the temperature range 2--300~K. The Hall mobility exceeds $10^{6}$~cm$^{2}$V$^{-1}$s$^{-1}$ at cryogenic temperatures and decreases systematically with increasing temperature, while the effective Hall carrier concentration exhibits strong carrier freeze-out behavior at low temperatures.

The combined evolution of Hall mobility, effective Hall carrier concentration, and resistivity reveals distinct transport regimes associated with carrier freeze-out, extrinsic conduction, and phonon-limited scattering. The transport behavior is interpreted within a Matthiessen’s-rule-inspired phenomenological mobility model, motivated by the combined influence of ionized impurity, neutral impurity, and acoustic phonon scattering. Variations among samples are correlated with differences in effective Hall carrier concentration and transport behavior.

These measurements establish a transport baseline for USD-grown high-resistivity germanium crystals and provide guidance for future material optimization toward detector-grade high-purity germanium for low-background rare-event detector applications.
\end{abstract}

\maketitle

\section{Introduction}

High-purity germanium (HPGe) is widely used in low-noise semiconductor radiation detectors for applications including $\gamma$-ray spectroscopy, neutrinoless double-beta decay searches, and direct dark-matter detection~\cite{avignone2008double, ABROSIMOV2020125396, osti_5184481, agostini2018improved, aalseth2011coherent}. Its performance is enabled by extremely low electrically active impurity concentration, high crystalline quality, and favorable band structure, which together allow long carrier drift lengths and excellent energy resolution at cryogenic temperatures~\cite{canali1975electron, jacoboni1983review}. As detector technologies advance toward sub-eV thresholds, a detailed understanding of charge transport over a broad temperature range is increasingly important \cite{cebrian2014transport, agostini2020charge}.

Charge transport in germanium has traditionally been described in terms of ionized impurity scattering at low temperatures and phonon scattering at higher temperatures, leading to characteristic temperature-dependent mobility behavior~\cite{brooks1955scattering, bardeen1950deformation, canali1975electron}. While this framework is adequate for moderately doped materials, it becomes incomplete in high-purity crystals, where multiple scattering mechanisms contribute simultaneously.

In this regime, neutral impurity scattering and incomplete dopant ionization play an increasingly important role, particularly at cryogenic temperatures. Carrier freeze-out further modifies the effective carrier density and scattering landscape, leading to complex, non-monotonic transport behavior~\cite{mei2016impact}. These effects cannot be fully captured by simplified two-component scattering models and require a more comprehensive description based on the combined contributions of multiple scattering processes \cite{Irisawa20031425, MYRONOV2025314}.

Despite prior studies, systematic experimental investigations of temperature-dependent transport in low impurity HPGe remain limited, particularly across the full 2--300~K range relevant for detector operation. In addition, the interplay between mobility, carrier concentration, and resistivity has not been consistently analyzed within a unified framework.

In this work, we present a comprehensive study of temperature-dependent charge transport in five lightly doped p-type HPGe samples grown at the University of South Dakota. Using combined Hall effect and four-probe resistivity measurements over 2--300~K, we analyze the evolution of mobility, effective Hall carrier concentration, and resistivity across distinct transport regimes. The results are interpreted within the Matthiessen’s-rule-inspired phenomenological mobility model, providing a consistent description of charge transport in HPGe crystals.

\section{Methods}
\subsection{Crystal Growth and Sample Preparation}

\begin{figure}[htbp]
    \centering
    \includegraphics[width=1.0\linewidth]{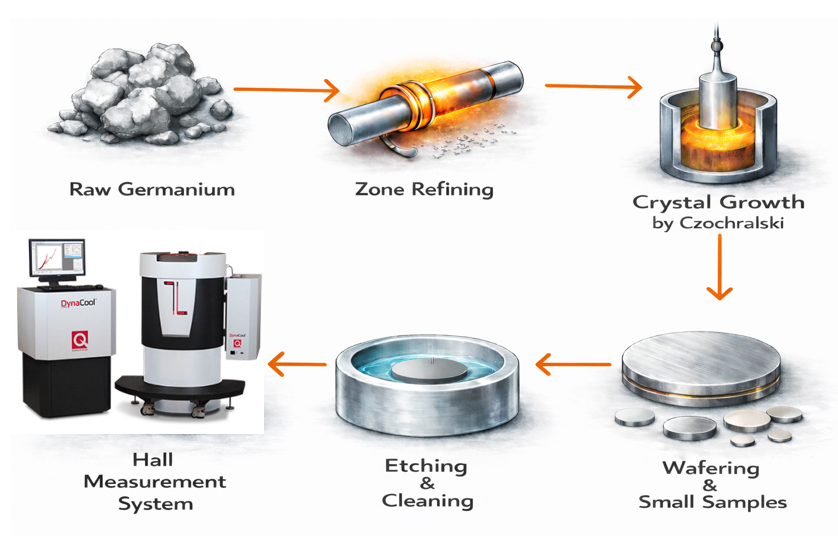}
    \caption{Schematic of the sample preparation workflow for Hall effect measurements, including zone refining, Czochralski crystal growth, wafering, polishing, chemical etching, and electrical characterization.}
    \label{fig:sample_prep}
\end{figure}

High-purity germanium (HPGe) crystals investigated in this work were grown at the University of South Dakota using zone refining followed by Czochralski (CZ) crystal growth. Multiple passes of zone refining were applied to the starting germanium feedstock to reduce electrically active impurities prior to crystal growth. The overall crystal growth and sample preparation process is illustrated in Fig.~\ref{fig:sample_prep}.

Single crystals were grown along the $\langle100\rangle$ crystallographic direction under a hydrogen atmosphere to suppress surface oxidation. A dash-necking procedure was employed during the initial growth stage to eliminate threading dislocations, followed by steady-state growth of the crystal boule~\cite{bhattarai2024investigating,budhathoki2025thickness}.

Sections of the grown crystal boules were sliced into wafers using a diamond wire saw and mechanically polished on both sides to obtain flat, parallel surfaces suitable for transport measurements. The resulting sample thicknesses were approximately $0.11$--$0.62~\mathrm{mm}$.

\begin{table*}[t]
\centering
\caption{Summary of sample geometry (length, width, and thickness) and room-temperature transport properties (resistivity $\rho_{300K}$, apparent Hall carrier concentration $p_H^{\mathrm{app}}(300\,\mathrm{K})$, and Hall mobility $\mu_{300K}$) for USD-grown HPGe samples. The quantity $p_H^{\mathrm{app}}(300\,\mathrm{K})$ denotes the apparent Hall carrier concentration extracted at room temperature using the single-carrier approximation. The quantity $p_H(77\,\mathrm{K})$ represents the effective Hall carrier concentration measured in the extrinsic transport regime, where intrinsic carrier contributions are comparatively small.}
\label{tab:samples}
\begin{tabular}{c c c c c c c c}
\hline\hline
Sample &
$L$ &
$W$ &
$t$ &
$\rho_{300K}$ &
$p_H^{\mathrm{app}}(300\,\mathrm{K})$ &
$p_H(77\,\mathrm{K})$ &
$\mu_{300K}$ \\
&
(mm) &
(mm) &
($\mu$m) &
($\Omega\cdot$cm) &
(cm$^{-3}$) &
(cm$^{-3}$) &
(cm$^{2}$/V$\cdot$s) \\
\hline

A1 & 2.21 & 3.27 & 118 &
43.45 &
$1.09\times10^{14}$ &
$3.59\times10^{12}$ &
$1.44\times10^{3}$ \\

A2 & 1.49 & 2.50 & 118 &
38.44 &
$1.14\times10^{14}$ &
$6.05\times10^{12}$ &
$1.38\times10^{3}$ \\

A3 & 1.70 & 1.15 & 118 &
42.17 &
$1.29\times10^{14}$ &
$4.14\times10^{12}$ &
$1.07\times10^{3}$ \\

B1 & 1.10 & 3.07 & 520 &
56.9 &
$8.43\times10^{13}$ &
$4.12\times10^{12}$ &
$1.56\times10^{3}$ \\

B2 & 1.00 & 2.79 & 620 &
58.3 &
$7.25\times10^{13}$ &
$3.97\times10^{12}$ &
$1.30\times10^{3}$ \\

\hline\hline
\end{tabular}
\end{table*}

A summary of the geometric parameters and transport properties for each sample is provided in Table~\ref{tab:samples}. The room-temperature values listed in Table~\ref{tab:samples} correspond to the apparent Hall carrier concentration, $p_H^{\mathrm{app}}(300~\mathrm{K})$, obtained using the single-carrier relation $p_H = 1/(e|R_H|)$. At elevated temperatures, intrinsic carrier generation and mixed electron--hole transport can influence the Hall coefficient, causing the extracted Hall carrier concentration to differ from the effective extrinsic carrier concentration. For comparison, Table~\ref{tab:samples} also lists $p_H(77~\mathrm{K})$, which corresponds to the effective Hall carrier concentration measured in the extrinsic regime and is more representative of the net ionized impurity concentration. Consequently, the room-temperature values should not be interpreted as the low-temperature extrinsic acceptor concentration discussed in Fig.~\ref{fig:T_n}.

The HPGe samples investigated in this work are high-resistivity, high-purity research-grade crystals and are not independently verified to meet detector-grade impurity specifications. Owing to Hall-factor corrections, compensation effects, and possible contributions from surface conduction and contact geometry, the Hall-derived carrier concentrations may differ from the true electrically active impurity concentration. Therefore, the carrier concentrations reported in this work are used primarily for comparative analysis between samples rather than as direct measures of absolute impurity density.

Prior to electrical measurements, samples were cleaned sequentially in acetone, isopropanol, and deionized water, followed by chemical etching in an HF:HNO$_3$ (1:3) solution to remove surface damage introduced during polishing for a few minutes. Ohmic contacts were subsequently formed at predefined positions for Hall effect and resistivity measurements. The contacts were verified to exhibit linear current–voltage characteristics, confirming ohmic behavior. Electrical contacts were positioned near the sample edges in a four-terminal configuration compatible with the PPMS Electrical Transport Option. The different contact arrangement was used for resistivity and Hall-effect measurements. Sample thickness, width, length, and voltage-probe spacing were measured using a digital Vernier caliper prior to mounting. The probe spacing (s) used in the resistivity calculations corresponds to the center-to-center distance between the voltage contacts.

Ohmic contacts were verified by room-temperature I–V measurements prior to cooldown. Hall measurements at all temperatures were performed using magnetic-field sweeps from (-14) T to (+14) T, and Hall coefficients were extracted from antisymmetrized data. Therefore, field-reversal checks were inherently included in all reported Hall measurements. No additional geometry correction factors beyond the dimensions were applied. Possible uncertainties associated with finite contact size, contact placement, and sample geometry are included within the overall systematic uncertainty estimates.

\subsection{Resistivity Measurement Using Four-Probe Technique}

\begin{figure}[htbp]
    \centering
    \includegraphics[width=1.0\linewidth]{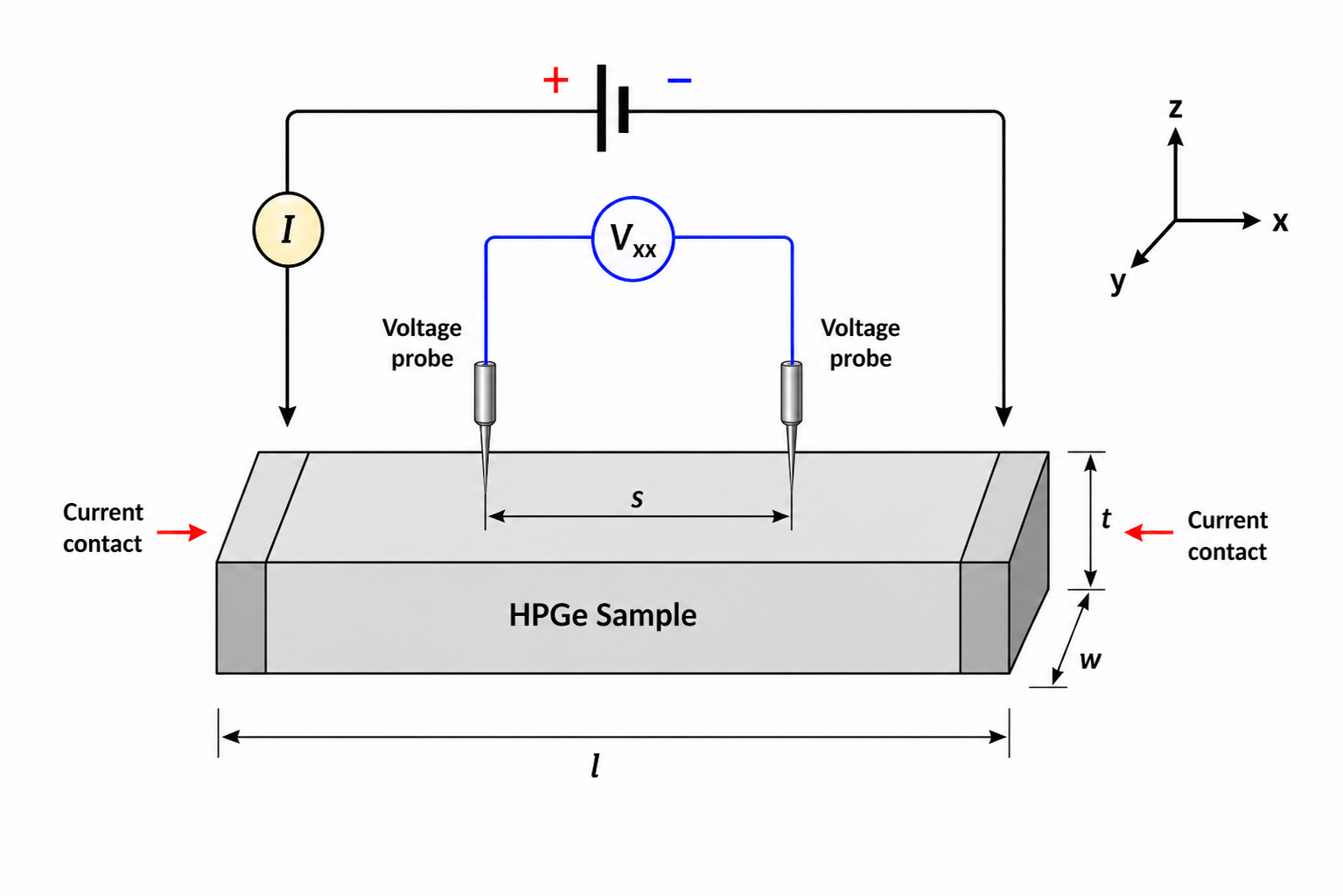}
    \caption{Schematic of the four-probe resistivity measurement configuration for an HPGe sample. A current $I$ is applied through the outer probes, while the longitudinal voltage drop $V_{xx}$ is measured between the inner voltage probes separated by a distance $s$. The sample dimensions are thickness $t$, width $w$, and length $l$.}

    \label{fig:resistivity_setup}
\end{figure}

The electrical resistivity of the HPGe samples was measured using a standard four-probe method, as illustrated in Fig.~\ref{fig:resistivity_setup}. This technique minimizes the contact resistance and ensures accurate extraction of the intrinsic bulk resistivity.

In this configuration, a constant current $I$ is injected through the outer two probes, while the voltage drop $V_{xx}$ is measured across the inner two probes. The probes are arranged linearly along the length of the sample, with a separation $s$ between them.

The longitudinal resistance $R_{xx}$ is obtained from Ohm's law as:
\begin{equation}
R_{xx} = \frac{V_{xx}}{I}.
\end{equation}

The resistivity $\rho_{xx}$ is then calculated by accounting for the sample geometry:
\begin{equation}
\rho_{xx} = R_{xx} \cdot \frac{Area}{Length} = R_{xx} \cdot \frac{t \cdot w}{s},
\end{equation}
where $t$ is the thickness and $w$ is the width of the sample.

The measurement assumes a uniform current distribution across the cross-sectional area and negligible edge effects, which is valid for samples where the probe spacing is small compared to the sample dimensions.

All measurements were performed under controlled temperature conditions ranging from 2~K to 300~K using a Physical Property Measurement System (PPMS), allowing extraction of temperature-dependent resistivity behavior of the HPGe samples.

\subsection{Hall Effect Measurement Procedure}

\begin{figure}[htbp]
    \centering
    \includegraphics[width=0.85\linewidth]{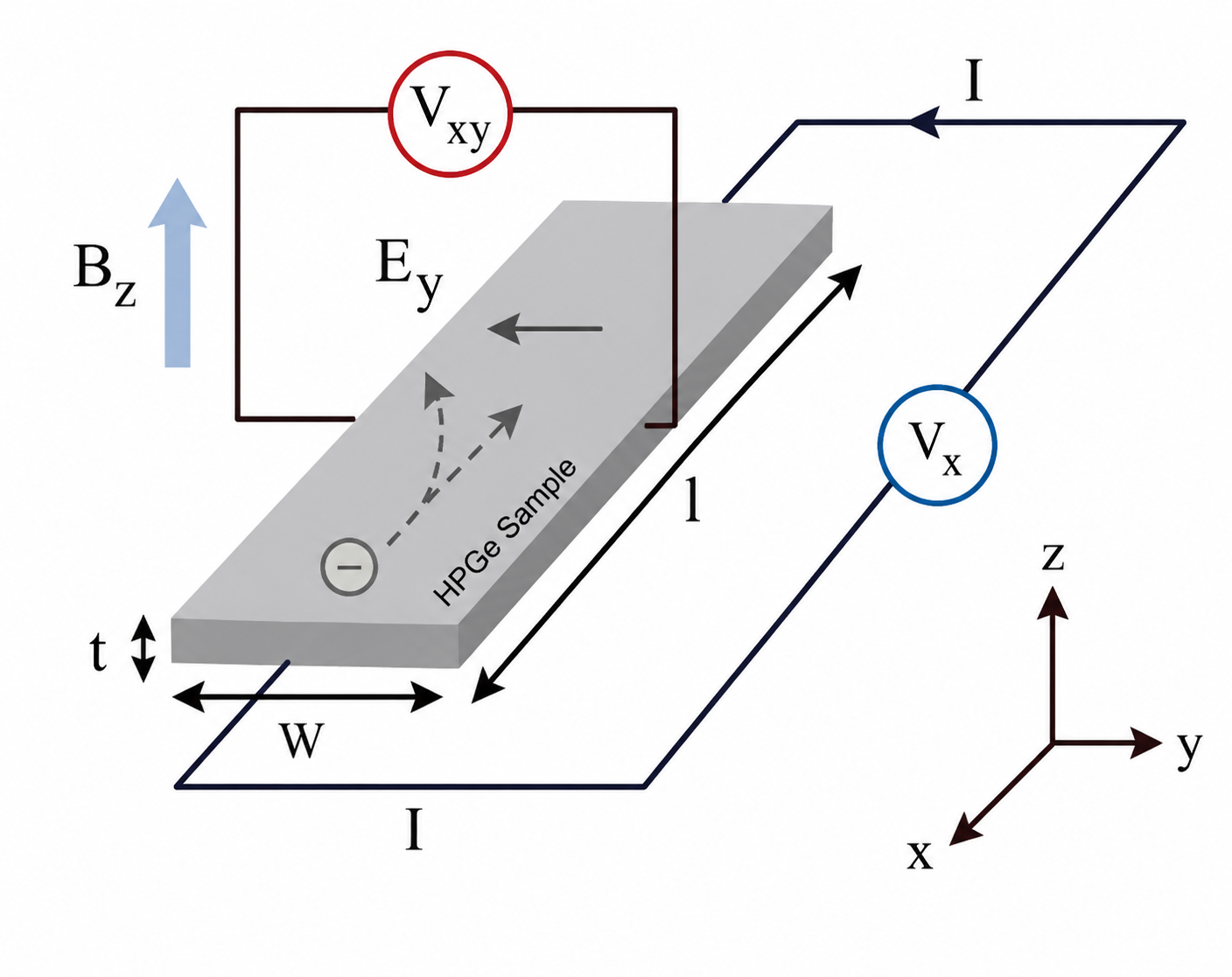}
    \caption{Schematic of the Hall-effect measurement configuration. A current $I$ is applied along the $x$-direction, while a magnetic field $B$ is applied along the $z$-direction, generating a transverse Hall voltage $V_{xy}$ along the $y$-direction.}
    \label{fig:hall}
\end{figure}

Hall effect measurements were performed over the temperature range 2--300~K using a Quantum Design PPMS DynaCool system equipped with the resistivity option. A standard four-probe Hall-bar configuration was employed, as illustrated in Fig.~\ref{fig:hall}.

In this geometry, a constant current $I$ is applied along the longitudinal ($x$) direction of the sample, while the transverse Hall voltage $V_{xy}$ is measured along the $y$ direction under an applied magnetic field $B$ perpendicular to the sample surface ($z$-direction). The Hall resistance is defined as
\begin{equation}
R_{xy} = \frac{V_{xy}}{I}.
\end{equation}
For a rectangular Hall-bar geometry with uniform current flow, the Hall resistivity depends upon the Hall resistance and thickness of the sample. In this configuration, no additional geometric correction factors (such as $width(w)/length(l)$) are required, as the Hall voltage is measured transverse to the current direction. The PPMS system provides the transverse Hall voltage $V_{xy}$, from which the Hall resistance is calculated. All reported values correspond to anti-symmetrized $R_{xy}(B)$ data, and no additional geometry correction beyond thickness scaling is applied.

At each temperature, the magnetic field was swept from $-14~\mathrm{T}$ to $+14~\mathrm{T}$. To eliminate contributions from longitudinal voltage due to contact misalignment, the Hall resistance was antisymmetrized according to \cite{abhi2024}
\begin{equation}
R_{xy}(B) = \frac{R(B) - R(-B)}{2}.
\label{eq:antisym}
\end{equation}

\begin{figure}[htbp]
\centering
\includegraphics[width=0.48\textwidth]{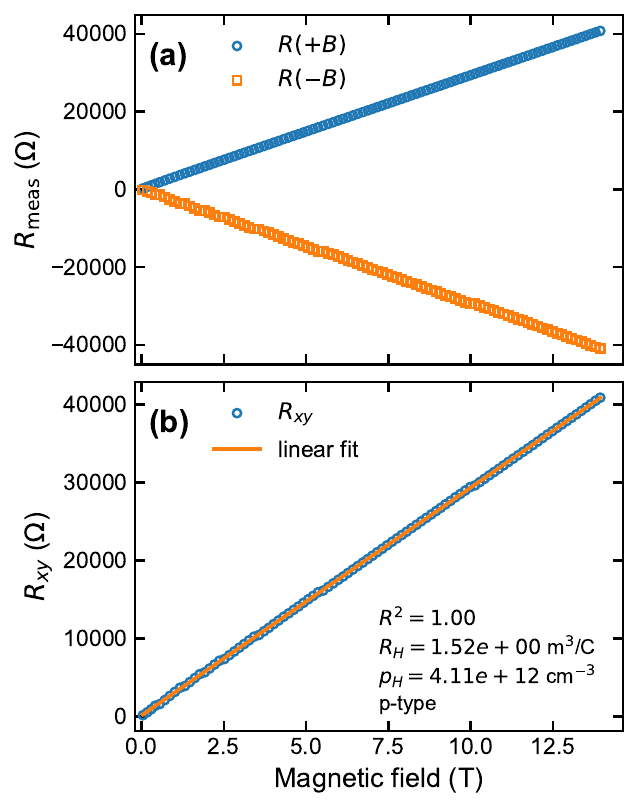}
\caption{Representative Hall-effect measurements for sample B1 at 77~K. (a) Raw Hall-channel resistance measured under positive and negative magnetic fields. The difference between the two field directions reflects the admixture of the longitudinal resistance contribution arising from contact misalignment and geometric imperfections. (b) Antisymmetrized Hall resistance ($R_{xy}(B)$) after removal of the symmetric longitudinal component. The solid line represents a linear fit used to determine the Hall coefficient and carrier concentration.}
\label{fig:hall_antisymmetrization}
\end{figure}

A representative example of the Hall-analysis procedure is
shown in Fig.~\ref{fig:hall_antisymmetrization}. The raw
Hall-channel resistance exhibits a small symmetric
contribution arising from contact misalignment and
geometric imperfections. After antisymmetrization, the
Hall resistance is nearly linear with magnetic field,
confirming dominant single-carrier transport and
demonstrating the reliability of the Hall-coefficient
extraction procedure.

The Hall resistivity is obtained from
\begin{equation}
\rho_{xy} = {R_{xy}\cdot t},
\end{equation}
where $t$ is the sample thickness. For a single dominant carrier type, $\rho_{xy}$ varies linearly with magnetic field,
\begin{equation}
\rho_{xy} = R_H B,
\end{equation}
where $R_H$ is the Hall coefficient extracted from the slope of $\rho_{xy}$ versus $B$.

For temperatures where transport is dominated by a single carrier type, the effective Hall carrier concentration is approximated by \cite{CHuang}
\begin{equation}
p_H = \frac{1}{e |R_H|}.
\end{equation}

At higher temperatures, where both electrons and holes may contribute simultaneously to transport, the Hall coefficient becomes a two-carrier quantity. In this mixed-transport regime, ${p_H}$ should therefore be interpreted as an apparent effective Hall carrier concentration rather than the true free-hole density. The Hall mobility is calculated as
\begin{equation}
\mu_H = \frac{|R_H|}{\rho_{xx}},
\end{equation}
where $\rho_{xx}$ is the longitudinal resistivity obtained independently from four-probe measurements.

It is noted that the extracted carrier concentration represents an apparent Hall carrier density and may differ from the true impurity concentration due to the Hall factor and its dependence on scattering mechanisms. All measurements were performed using low excitation currents to minimize Joule heating effects.

The uncertainty in the measured transport quantities arises from both instrumental and geometric contributions. The PPMS DynaCool Electrical Transport Option employs a Kelvin four-probe configuration, for which the intrinsic electronics uncertainty is comparatively small. In the present work, the dominant uncertainty originates from sample geometry and contact definition, including thickness nonuniformity, probe spacing, lateral dimensions, contact alignment, Hall-slope fitting, and measurement repeatability.

The resistivity uncertainty is primarily determined by uncertainties in sample thickness, probe spacing, and resistance measurement, while the Hall coefficient uncertainty reflects contributions from Hall-voltage determination, Hall-slope fitting, thickness uncertainty, and residual contact misalignment. Representative uncertainty propagation using measured geometric tolerances and fitting uncertainties yields approximately 7 {\%} uncertainty for both resistivity and effective Hall carrier concentration. Because the Hall mobility depends on both the Hall coefficient and resistivity, its propagated uncertainty is approximately 10 {\%}.

These values represent conservative, representative uncertainties applied across the full temperature range rather than point-by-point statistical uncertainties. At intermediate temperatures, where carrier freeze-out, mixed transport behavior, and scattering transitions occur simultaneously, the actual uncertainty may be somewhat larger.

\subsection{Mobility Modeling and Scattering Mechanisms}

The temperature dependence of the Hall mobility was analyzed within the Matthiessen’s-rule-inspired phenomenological mobility model appropriate for high-purity germanium. In general, charge transport in semiconductors is influenced by several scattering mechanisms, and the total mobility can be expressed using Matthiessen’s rule \cite{jacoboni1983review},

\begin{equation}
\frac{1}{\mu(T)} =
\frac{1}{\mu_{\mathrm{I}}(T)} +
\frac{1}{\mu_{\mathrm{N}}(T)} +
\frac{1}{\mu_{\mathrm{ac}}(T)} +
\frac{1}{\mu_{\mathrm{op}}(T)} +
\frac{1}{\mu_{\mathrm{dis}}(T)}
\label{eq:mathessien_complete_equation}
\end{equation}

Here, $\mu_{\mathrm{I}}$, $\mu_{\mathrm{N}}$, $\mu_{\mathrm{ac}}$, $\mu_{\mathrm{op}}$, and $\mu_{\mathrm{dis}}$ denote the contributions from ionized impurity, neutral impurity, acoustic phonon, optical phonon, and dislocation scattering, respectively.

For HPGe crystals, optical phonon scattering is expected to remain weaker than acoustic phonon scattering over the range 2--300~K~\cite{brown1962analysis}, and dislocation scattering is expected to be minimal due to the low dislocation density achieved through crystal growth and necking procedures \cite{mei2016impact}. Therefore, the dominant contributions to charge transport can be approximated as

\begin{equation}
\frac{1}{\mu(T)} =
\frac{1}{\mu_{\mathrm{I}}(T)} +
\frac{1}{\mu_{\mathrm{N}}(T)} +
\frac{1}{\mu_{\mathrm{ac}}(T)}.
\label{eq:reduced_model}
\end{equation}

The ionized impurity mobility is modeled using a two-band formulation accounting for light-hole and heavy-hole contributions \cite{mei2016impact},

\begin{equation}
\mu_{\mathrm{I}}(T) =
\frac{\mu_{\ell}(T) + 16.05\,\mu_{h}(T)}{17.05},
\end{equation}

where

\begin{equation}
\mu_{\ell}(T) =
\frac{4.021 \times 10^{18}}{N_i}
\frac{T^{3/2}}{\ln\!\left( \dfrac{9.125 \times 10^{13} T^2}{N_i} \right)},
\end{equation}

\begin{equation}
\mu_{h}(T) =
\frac{1.594 \times 10^{18}}{N_i}
\frac{T^{3/2}}{\ln\!\left( \dfrac{5.8 \times 10^{14} T^2}{N_i} \right)},
\end{equation}

and $N_i$ is the ionized impurity concentration.

The neutral impurity scattering contribution is given by \cite{mcgill1975neutral, ghosh2019impurity, palleti2024properties}

\begin{equation}
\mu_{\mathrm{N}}(T) =
9.76 \times 10^{4}
\left( 0.228\,T^{1/2} + 0.976\,T^{-1/2} \right),
\end{equation}

while the acoustic phonon contribution follows

\begin{equation}
\mu_{\mathrm{ac}}(T) =
3.37 \times 10^{7}\,T^{-3/2}.
\end{equation}

At low temperatures, incomplete dopant ionization (carrier freeze-out) modifies both the effective Hall carrier concentration and effective scattering rates, leading to deviations from idealized transport behavior \cite{look1999mobility, conwell1950scattering}. To capture the dominant temperature dependence of the measured mobility, we employ an Matthiessen’s-rule-inspired phenomenological mobility model,

\begin{equation}
\frac{1}{\mu_H(T)} = C\,T^{3/2} + D\,T^{-3/2},
\label{eq:empirical_model}
\end{equation}

where the $T^{3/2}$ term reflects the temperature dependence associated with phonon-limited transport, while the $T^{-3/2}$ term captures impurity-related and weakly temperature-dependent scattering contributions. This representation should be interpreted strictly as a phenomenological description of the overall temperature dependence of the mobility rather than as a unique decomposition of microscopic scattering mechanisms. Owing to the limited number of experimental temperature points available for each sample, the fitted parameters are intended primarily as compact descriptors of the observed transport trends and should not be interpreted as quantitative measures of individual ionized-impurity, neutral-impurity, or phonon-scattering contributions.

The fitting parameters $C$ and $D$ have units of
$\mathrm{V\,s\,cm^{-2}\,K^{-3/2}}$
and
$\mathrm{V\,s\,cm^{-2}\,K^{3/2}}$,
respectively, and were extracted using least-squares fitting over the temperature ranges where the Hall signal remained reliable. These parameters provide a comparative measure of effective scattering strength and material quality among the samples.

\section{Results}
\subsection{Resistivity Measurement}
\begin{figure}[htbp]
    \centering
    \includegraphics[width=0.5\textwidth]{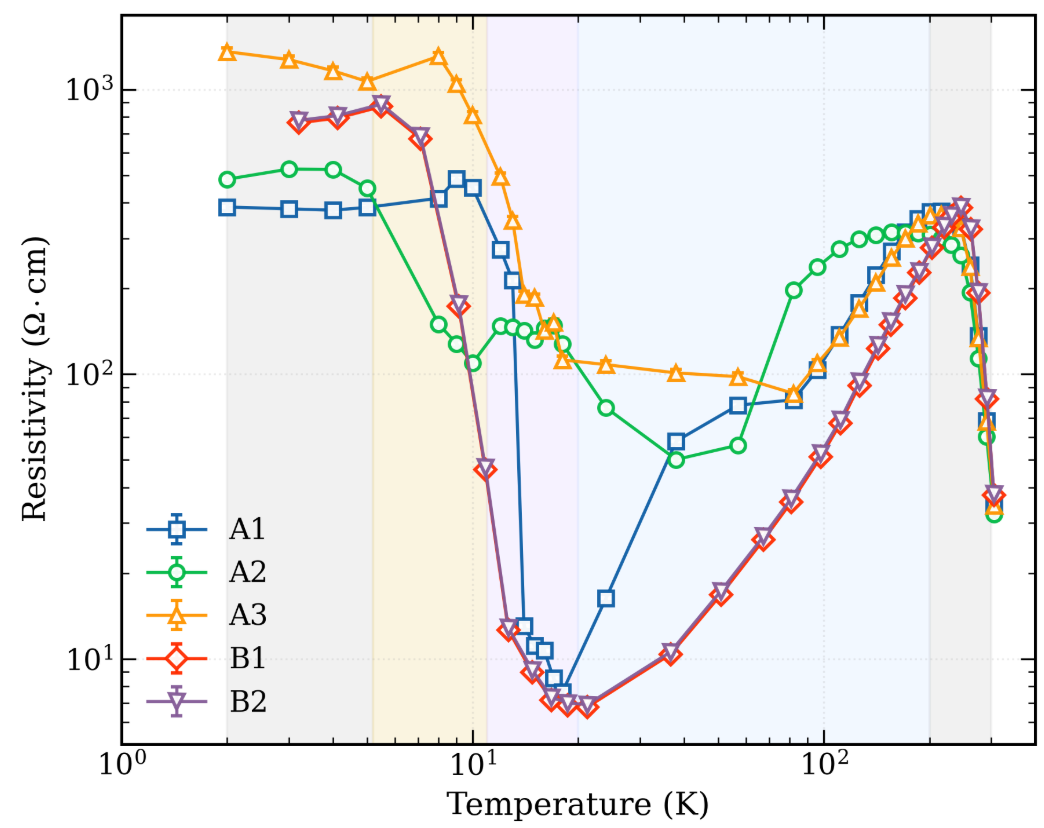}
\caption{Temperature dependence of the electrical resistivity for USD-grown p-type HPGe samples. Symbols represent measured data points for samples A1--A3 and B1--B2. Error bars correspond to a representative 7\% uncertainty, primarily arising from geometric factors and contact definition. The shaded background regions indicate approximate transport regimes: localized transport ($T \lesssim 5.2$ K), freeze-out transition ($\sim 5.2$--11 K), impurity-ionization crossover ($\sim 11$--20 K), extrinsic mobility-limited transport ($\sim 20$--200 K), and high-temperature mixed transport ($T \gtrsim 200$ K).}
    \label{fig:rhoT}
\end{figure}

Figure~\ref{fig:rhoT} presents the measured temperature dependence of the electrical resistivity $\rho(T)$ for USD-grown HPGe samples (A1–A3, B1–B2). All samples exhibit a strongly non-monotonic behavior spanning nearly two orders of magnitude in resistivity, characteristic of high-purity germanium. This behavior reflects the interplay between carrier freeze-out, impurity ionization, mobility-limited transport, and intrinsic carrier generation.

\vspace{0.3cm}
\noindent
\textbf{Regime I: Localized transport ($T \lesssim 5.2$ K).}

At the lowest temperatures, the resistivity is high 
($\rho \sim 2\times10^2$–$10^3~\Omega\cdot\text{cm}$) and exhibits a weak 
temperature dependence across all samples. This behavior is consistent with a substantial fraction of carriers being frozen into localized impurity states. The observed temperature dependence suggests strongly localized carrier behavior and possible hopping-like transport within impurity-related localized states, although the presently available data do not permit a definitive identification of the underlying conduction mechanism.

In this regime, the resistivity may be described by several hopping-transport models that have been applied previously to localized semiconductor systems, including nearest-neighbor hopping (NNH) and variable-range hopping (VRH)~\cite{Mott1969, EfrosShklovskii1975}.

For NNH, the resistivity 
follows an activated form,
\begin{equation}
    \rho(T) \propto \exp\!\left(\frac{W}{k_B T}\right), 
\end{equation}
where \(W\) represents the characteristic hopping activation energy associated with carrier transport between localized impurity states, and \(k_B\) is the Boltzmann constant.
while VRH conduction is described by
\begin{equation}
    \rho(T) \propto \exp\!\left[\left(\frac{T_0}{T}\right)^\alpha\right],
\end{equation}
with $\alpha = 1/4$ for Mott variable-range hopping in three dimensions~\cite{Mott1969} 
and $\alpha = 1/2$ for Efros--Shklovskii hopping in the presence of a 
Coulomb gap~\cite{EfrosShklovskii1975}. Here, $T_0$ is a characteristic temperature that depends on the density of localized states near the Fermi level and the carrier localization length, and reflects the degree of disorder and localization in the system.

The present data exhibit behavior qualitatively consistent with hopping-like transport; however, the presently available temperature sampling does not permit a quantitative distinction among activated transport, nearest-neighbor hopping, variable-range hopping, or Efros--Shklovskii-type conduction models.

This behavior is consistent with previous studies of high-purity 
germanium, such as the work by Mei~\cite{Mei2024ResidualImpurities}, 
where the majority of carriers are bound to acceptor states at low 
temperature. For p-type material, the fraction of bound holes can be 
expressed as
\begin{equation}
    f_{b,A}(T) \equiv \frac{p_a}{p + p_a}
    =
    \left[
    1 + \frac{N_V(T)}{4N_A}
    \exp\!\left(-\frac{E_A - E_V}{k_B T}\right)
    \right]^{-1}
    \label{fraction_p},
\end{equation}
where $p_a$ is the number of holes bound to acceptors, $p$ is the free 
hole concentration, $N_V$ is the effective density of states in the 
valence band, and $E_A - E_V$ is the acceptor binding energy. The factor 
of $1/4$ arises from the degeneracy of the valence band (light and heavy 
hole bands with spin states). A quantitative extraction of the acceptor activation energy from Arrhenius analysis was not performed due to the limited number of data points in the freeze-out regime. However, the observed temperature range is consistent with shallow acceptor levels in germanium ($\sim 10$ meV).

As a result, the free carrier concentration is strongly suppressed in 
this temperature range, and the electrical transport is consistent with carrier localization and freeze-out effects, with possible contributions from hopping-like transport between impurity-related localized states.

\vspace{0.3cm}
\noindent
\textbf{Regime II: Freeze-out transition ($T \sim 5.2$–$11$ K).}

A gradual decrease in resistivity is observed between approximately 
$T \sim 5$ K and $T \sim 11$ K across all samples, where $\rho(T)$ 
drops by nearly two orders of magnitude. This behavior corresponds 
to the freeze-out transition, during which carriers are thermally 
ionized from impurity-bound states into the valence band.

Within the framework developed by Mei~\cite{Mei2024ResidualImpurities}, 
this process is described in terms of the temperature-dependent 
fraction of holes bound to acceptor states. For p-type material, 
the bound fraction is given by (Eq. ~\ref{fraction_p}).

The free carrier fraction is then given by
\begin{equation}
    f_{\mathrm{free}}(T) = 1 - f_{b,A}(T),
\end{equation}
so that the hole concentration can be approximated as
\begin{equation}
    p(T) \simeq N_A \, f_{\mathrm{free}}(T).
\end{equation}

In this temperature interval, the rapid increase of 
$f_{\mathrm{free}}(T)$ dominates the transport behavior, leading 
to a strong reduction in resistivity. The resistivity can therefore 
be expressed as
\begin{equation}
    \rho(T) = \frac{1}{q \, \mu(T) \, N_{\mathrm{eff}} \, f_{\mathrm{free}}(T)},
\end{equation}
where $N_{\mathrm{eff}} \simeq |N_A - N_D|$ is the effective dopant 
concentration.

Because the carrier density changes exponentially with temperature 
in this regime, the dominant contribution to the temperature dependence 
of $\rho(T)$ arises from impurity ionization, while the mobility 
variation $\mu(T)$ plays a comparatively minor role. The observed decrease in resistivity is therefore a direct manifestation of 
the thermal activation of carriers from shallow acceptor levels.

\vspace{0.3cm}
\noindent
\textbf{Regime III: Impurity ionization crossover ($T \sim 11$–$20$ K).}

Following the freeze-out transition, the resistivity reaches a minimum 
in the range $T \sim 11$–$20$ K, with $\rho \sim 5$–$15~\Omega\cdot\text{cm}$. 
This minimum arises from the competition between the increasing free 
carrier concentration due to impurity ionization and the onset of 
mobility reduction with increasing temperature.

In this regime, a substantial fraction of dopants becomes ionized, 
and transport begins to transition from localized impurity-band 
conduction toward band-like transport. The effective carrier concentration 
in the nondegenerate limit can be approximated as
\begin{equation}
    p(T) \propto T^{3/2} \exp\!\left(-\frac{\Delta}{k_B T}\right),
\end{equation}
where $\Delta = E_A - E_V$ is the acceptor activation energy.

Although the carrier density increases rapidly with temperature, 
the mobility begins to decrease due to enhanced scattering, leading 
to the observed minimum in $\rho(T)$. The position and depth of this 
minimum vary between samples, reflecting differences in impurity 
concentration, compensation level, and crystal quality.

\vspace{0.3cm}
\noindent
\textbf{Regime IV: Extrinsic, mobility-limited transport ($T \sim 20$–$200$ K).}

In the intermediate temperature range, the resistivity increases monotonically with temperature for all samples. This behavior indicates that the extrinsic Hall carrier concentration is approximately saturated, and that transport is limited by carrier mobility. In this regime, mobility is dominated by acoustic phonon scattering~\cite{sze2007chapter},
\begin{equation}
    \mu(T) \propto T^{-m}, \quad m \approx 1.5,
\end{equation}
leading to an increase in resistivity with temperature. The observed trend is consistent with Matthiessen’s rule, where phonon scattering dominates over impurity scattering in this temperature range.

\vspace{0.3cm}
\noindent
\textbf{Regime V: High-temperature mixed transport ($T \gtrsim 200$ K).}

At higher temperatures, the resistivity reaches a maximum and subsequently decreases with increasing temperature. This turnover reflects the onset of a high-temperature mixed transport regime, where thermally generated intrinsic carriers begin to contribute to the conductivity in addition to impurity-derived carriers. 

For germanium, the intrinsic carrier concentration at room temperature is approximately $n_i(300~\mathrm{K}) \sim 2 \times 10^{13}~\mathrm{cm^{-3}}$, which is comparable to or smaller than the room-temperature effective Hall carrier concentration measured in the present samples. Therefore, the system does not reach a purely intrinsic regime within the measured temperature range. Instead, the observed behavior arises from the combined effects of phonon-limited mobility, impurity-derived carriers, and partial intrinsic carrier generation. In this temperature regime, both electrons and holes may contribute to the Hall response. Consequently, the relation $p_H = 1/(e|R_H|)$ no longer represents a true single-carrier hole concentration, and the extracted Hall carrier density should be interpreted qualitatively as an effective transport parameter.

The intrinsic carrier concentration follows
\begin{equation}
    n_i(T) \propto T^{3/2} \exp\!\left(-\frac{E_g}{2k_B T}\right),
\end{equation}
and increases rapidly with temperature. Although intrinsic carriers do not dominate in the present samples, their growing contribution at high temperatures leads to enhanced conductivity despite continued mobility reduction due to phonon scattering.
\vspace{0.3cm}
\noindent

\textbf{Sample-to-sample variations.}

While all samples exhibit the same qualitative transport behavior, quantitative differences are observed. Samples A3 and B2 show higher resistivity at low temperature, indicating stronger localization or lower effective carrier density, whereas samples such as A1 exhibit a sharper freeze-out transition and lower resistivity minimum, suggesting higher dopant activation or reduced compensation. These variations are consistent with differences in impurity concentration, defect density, and crystal quality across the samples.
The measured $\rho(T)$ behavior demonstrates a clear progression from transport consistent with carrier localization, with possible hopping-like contributions at low temperatures, through impurity ionization and mobility-limited transport, to mixed transport with increasing intrinsic carrier contributions at high temperatures. The experimentally observed transition temperatures and resistivity scales are consistent with expectations for high-purity germanium and provide a quantitative basis for understanding charge transport in these HPGe crystals.

\subsection{Hall carrier concentration}

\begin{figure}[htbp]
    \centering
    \includegraphics[width=0.5\textwidth]{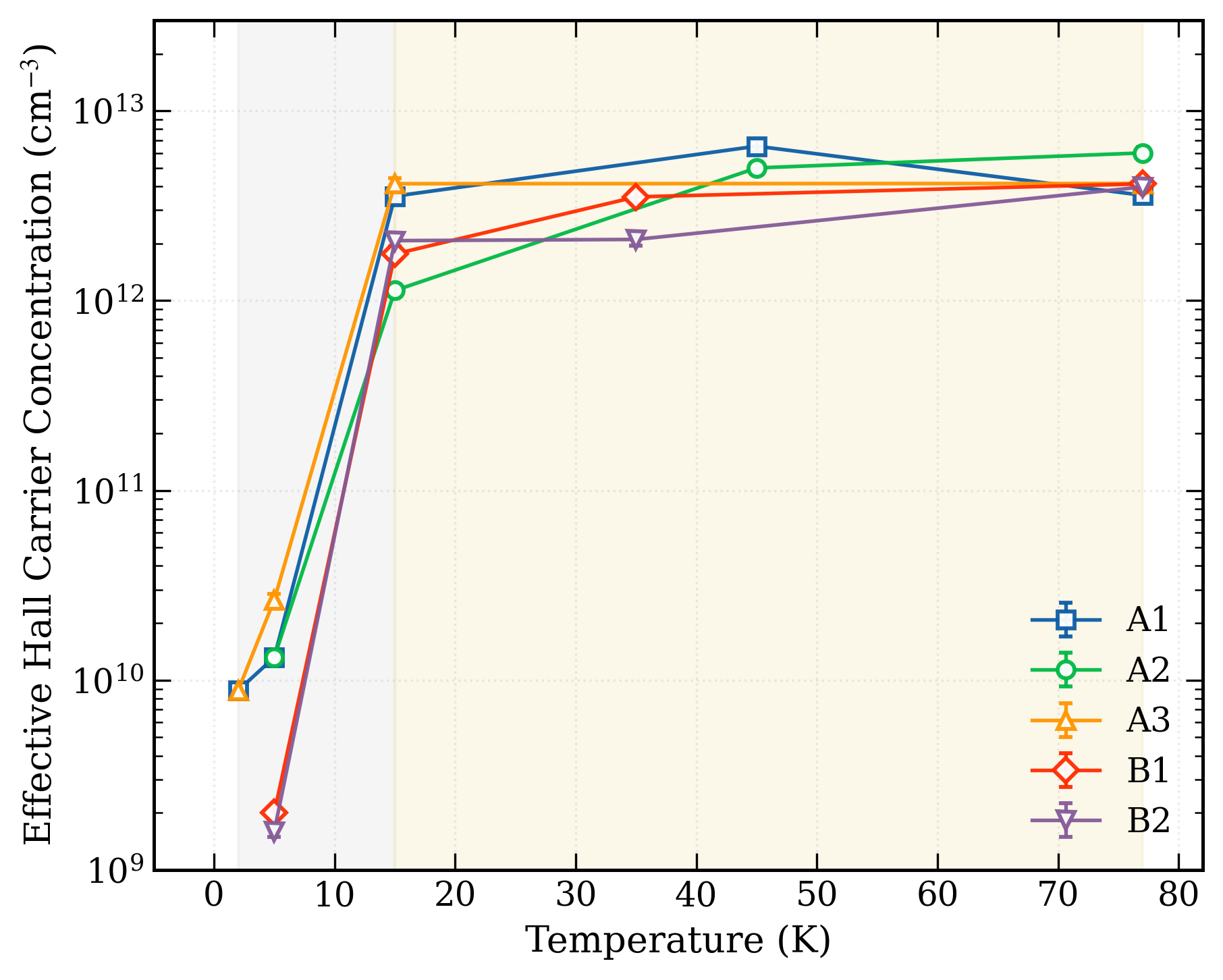}
    \caption{Temperature dependence of the effective Hall carrier concentration for USD-grown p-type HPGe samples. The shaded regions indicate the low-temperature carrier freeze-out regime and the higher-temperature apparent extrinsic Hall plateau. Data above 80~K are omitted because intrinsic carrier contributions become significant. Error bars correspond to a representative 7\% uncertainty.}
    \label{fig:T_n}
\end{figure}

Figure~\ref{fig:T_n} shows the temperature dependence of the effective Hall carrier concentration, $p_H = 1/(e|R_H|)$, for the USD-grown HPGe samples. The temperature axis is divided into shaded regions to highlight the transition from the low-temperature freeze-out regime to the higher-temperature extrinsic regime. It is important to note that $p_H$ represents an effective carrier density obtained from Hall measurements and may differ from the true ionized impurity concentration due to Hall factor effects and scattering-dependent corrections.

At temperatures above $\sim 15$ K, corresponding to the right shaded region, the extrinsic Hall carrier concentration exhibits a weak temperature dependence and approaches a plateau with values on the order of 3 $\times$$10^{12}~\mathrm{cm^{-3}}$. This behavior corresponds to the extrinsic regime, where most acceptor impurities are ionized. In this regime, $p_H$ provides an approximate comparative estimate of the effective ionized acceptor density, allowing relative differences between samples to be assessed.

As the temperature decreases below $\sim 15$ K, corresponding to the left shaded region, the onset of freeze-out begins and becomes progressively more pronounced with decreasing temperature. The value of $p_H$ decreases rapidly by several orders of magnitude, indicating incomplete dopant ionization and the onset of carrier freeze-out. In this regime, the strong reduction in carrier density dominates transport, leading to increased resistivity despite the concurrent increase in mobility.

At the lowest temperatures ($T \lesssim 5~\mathrm{K}$), the effective Hall carrier concentration falls below $10^{10}~\mathrm{cm^{-3}}$ for the highest-purity samples, confirming deep freeze-out of acceptor states. The persistence of a measurable Hall signal suggests transport mediated by a small population of thermally activated or impurity-band carriers.

Significant variations are observed among samples in both the magnitude of the high-temperature plateau and the sharpness of the freeze-out transition. In particular, samples B1 and B2 exhibit lower effective carrier concentrations, consistent with lower effective impurity levels inferred from Hall measurements. These differences correlate with the observed variations in resistivity (Fig.~\ref{fig:rhoT}) and mobility (Fig.~\ref{fig:mu_T}), demonstrating that charge transport in HPGe is governed by the interplay between temperature-dependent carrier ionization and scattering mechanisms. Data above 80~K are not shown in Fig.~\ref{fig:T_n} because, at higher temperatures, intrinsic carrier generation increasingly influences the Hall coefficient, leading to a mixed electron--hole transport response. Consequently, the quantity $p_H = 1/(e|R_H|)$ should be regarded as an apparent Hall carrier concentration rather than a direct measure of the free-hole concentration.

\subsection{Temperature-dependent Hall mobility}

\begin{figure}[htbp]
    \centering
    \includegraphics[width=0.5\textwidth]{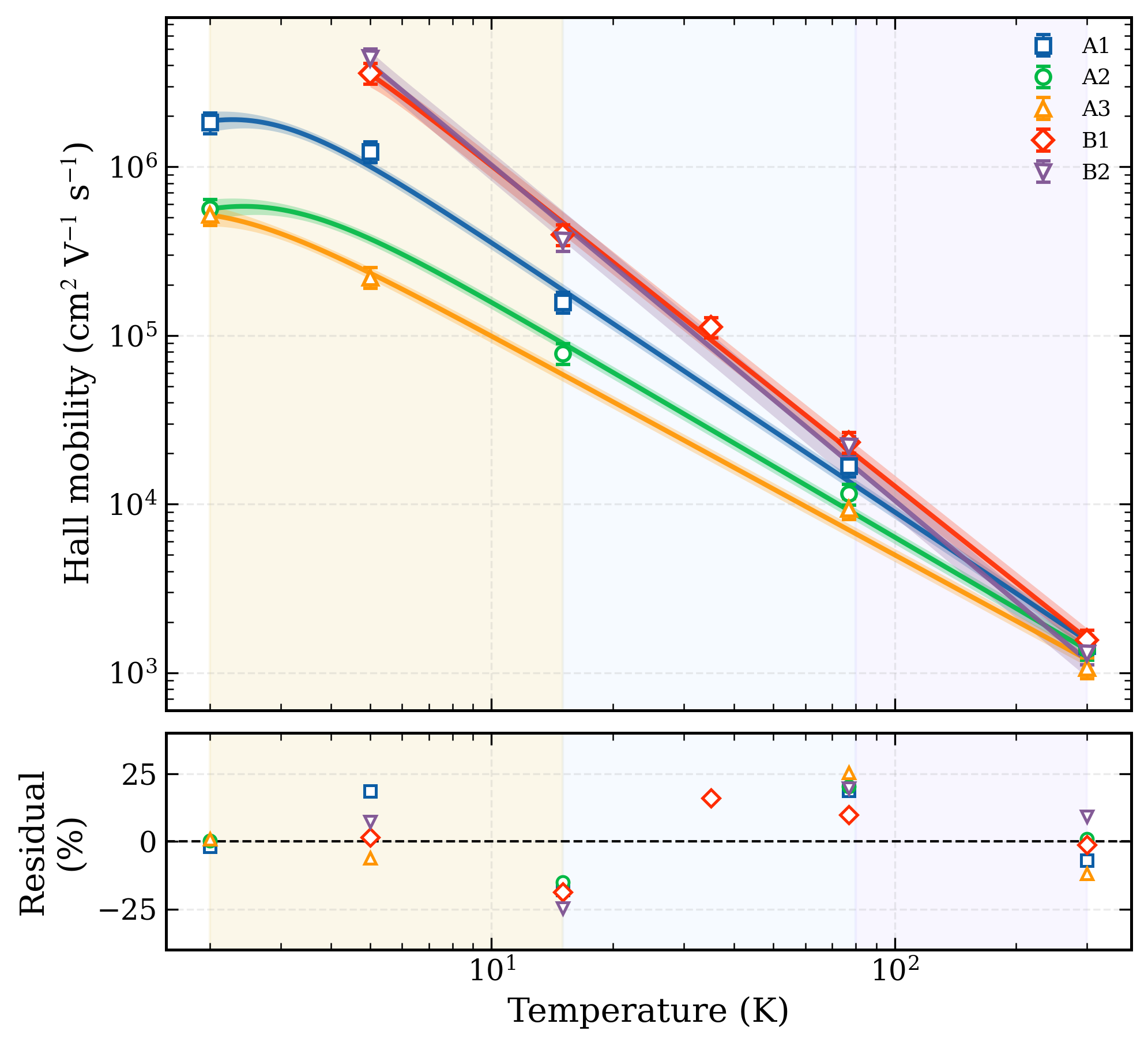}
    \caption{Temperature dependence of the Hall mobility for USD-grown high-resistivity germanium samples together with phenomenological multi-scattering fits based on Matthiessen's-rule-inspired transport behavior. Symbols represent experimental measurements, while solid curves show the fitted mobility model. Error bars correspond to the estimated experimental uncertainty (approximately 10\%), and the shaded regions around the fitted curves represent the propagated $1\sigma$ fitting uncertainty. The lower panel shows the percentage residuals between the measured data and fitted model. Shaded background regions indicate the approximate low-temperature freeze-out/carrier-activation regime ($T \lesssim 15$ K), intermediate mixed-scattering regime ($\sim 15$--80 K), and high-temperature phonon-dominated transport regime ($T \gtrsim 80$ K).}
    \label{fig:mu_T}
\end{figure}

Figure~\ref{fig:mu_T} shows the temperature dependence of the Hall mobility $\mu_H$ for five lightly doped p-type HPGe samples over the range $2$--$300~\mathrm{K}$. The mobility increases strongly with decreasing temperature, spanning more than two orders of magnitude from $\sim 10^{3}$~cm$^{2}$V$^{-1}$s$^{-1}$ at room temperature to above $10^{6}$~cm$^{2}$V$^{-1}$s$^{-1}$ at cryogenic temperatures. The experimental data (symbols) are reasonably well reproduced by the fitted curves (solid lines) within the selected fitting range, indicating that the adopted Matthiessen's-rule-inspired phenomenological mobility model captures the dominant transport trends. The effective model in (Eq.~\ref{eq:empirical_model}) provides a phenomenological description of the dominant temperature dependence of the mobility. While it captures the overall behavior, it does not allow a unique separation of individual microscopic scattering contributions, and therefore specific mechanisms should be interpreted qualitatively rather than quantitatively. To guide interpretation, the temperature axis is divided into shaded regions indicating approximate transport regimes, as discussed below.

At high temperatures ($T \gtrsim 80$--$100~\mathrm{K}$), corresponding to the rightmost shaded region, the mobility values for all samples converge and follow a similar temperature dependence, consistent with phonon-limited transport. In this regime, the mobility is broadly consistent with a power-law behavior $\mu \propto T^{-3/2}$, indicating that acoustic phonon scattering dominates and impurity-related contributions are comparatively weak. This behavior is consistent with the increase in resistivity observed in Fig.~\ref{fig:rhoT} despite nearly temperature-independent carrier concentration in Fig.~\ref{fig:T_n}.

In the intermediate temperature range ($\sim15$--$80~\mathrm{K}$), highlighted by the central shaded region, deviations from the high-temperature scaling become evident and the mobility begins to separate between samples. This separation reflects the increasing influence of impurity-related scattering as phonon scattering weakens. The observed trends correlate with the effective Hall carrier concentration extracted from Hall measurements (Fig.~\ref{fig:T_n}), where samples with lower plateau $p_H$ generally exhibit higher mobility, suggesting reduced effective impurity scattering.

At low temperatures ($T \lesssim 15~\mathrm{K}$), corresponding to the leftmost shaded region, the mobility increases rapidly with decreasing temperature but does not exhibit clear saturation within the measured range. This behavior indicates that impurity-related scattering remains relevant in the cryogenic regime. In high-purity germanium, neutral impurity scattering is expected to play an important role, while ionized impurity scattering introduces additional sample-dependent variation. Furthermore, carrier freeze-out, evidenced by the strong reduction in $p_H$ (Fig.~\ref{fig:T_n}), modifies the effective carrier population and complicates direct interpretation of the Hall mobility in this regime. At these temperatures, $\mu_H$ should therefore be interpreted as an apparent Hall mobility rather than the true drift mobility.

The experimental data are analyzed using an Matthiessen’s-rule-inspired phenomenological mobility model, in which the inverse mobility is expressed as a sum of temperature-dependent and weakly temperature-dependent contributions. The fitted parameters obtained from this model are summarized in Table~\ref{tab:fit}. The fits reproduce the overall temperature dependence of the mobility, particularly in the intermediate and high-temperature regimes, but deviations are observed at the lowest temperatures where carrier freeze-out and additional scattering mechanisms are not fully captured by the simplified model.

The fitted parameters show that both $C$ and $D$ contribute to the mobility behavior over the selected temperature range. For most samples, the uncertainty in $D$ is smaller than or comparable to its fitted value, indicating that the impurity-related scattering term is not negligible. However, because the fit is performed over a limited number of temperature points, the extracted parameters should be interpreted phenomenologically rather than as unique microscopic scattering constants. The variation of $C$ and $D$ among samples suggests differences in impurity concentration, defect structure, and carrier scattering mechanisms. It should be emphasized that the mobility model is intended primarily as a compact phenomenological description of the measured temperature dependence. The limited number of temperature points available for each sample does not permit a statistically robust separation of ionized-impurity, neutral-impurity, and phonon-scattering contributions. Consequently, the extracted parameters should be interpreted as trend descriptors rather than unique microscopic scattering constants.

Significant sample-to-sample variations are observed, particularly at low temperatures. Samples B1 and B2 exhibit the highest mobilities, suggesting lower effective impurity scattering and improved material quality. In contrast, sample A3 shows consistently lower mobility across the measured range, consistent with a higher effective impurity level or increased compensation.

\begin{table*}[!t]
\centering
\footnotesize
\setlength{\tabcolsep}{6pt}

\caption{Fitted parameters for the effective phenomenological mobility model obtained from least-squares fitting to the Hall mobility data. The fit range, number of data points ($N$), and reduced chi-square values are also listed. The parameters $C$ and $D$ should be interpreted as effective phenomenological coefficients that provide a compact description of the measured mobility trends. Because the fits are based on a limited number of temperature points, the extracted parameters are not intended to represent unique microscopic scattering contributions. Uncertainties correspond to the standard errors obtained from the covariance matrix.}

\begin{tabular}{c c c c c c c c}
\hline\hline
Sample & Fit Range (K) & $N$ & $C$ & $\sigma_C$ & $D$ & $\sigma_D$ & $\chi^2_\nu$ \\
\hline

A1 & 2--300 & 5
& $7.03\times10^{-8}$
& $6.21\times10^{-9}$
& $9.81\times10^{-7}$
& $2.33\times10^{-7}$
& 1.83 \\

A2 & 2--300 & 4
& $2.49\times10^{-7}$
& $2.16\times10^{-8}$
& $3.01\times10^{-6}$
& $6.88\times10^{-7}$
& 1.62 \\

A3 & 2--300 & 4
& $5.03\times10^{-7}$
& $4.45\times10^{-8}$
& $1.83\times10^{-6}$
& $7.91\times10^{-7}$
& 2.11 \\

B1 & 5--300 & 5
& $1.24\times10^{-8}$
& $1.93\times10^{-9}$
& $4.02\times10^{-7}$
& $9.55\times10^{-8}$
& 1.20 \\

B2 & 5--300 & 4
& $9.97\times10^{-9}$
& $2.02\times10^{-9}$
& $1.17\times10^{-8}$
& $1.04\times10^{-9}$
& 2.91 \\

\hline\hline
\end{tabular}

\label{tab:fit}
\end{table*}

Overall, the evolution of $\mu_H(T)$ demonstrates a continuous transition from phonon-limited transport at high temperatures to impurity-influenced transport at low temperatures, with additional modification due to carrier freeze-out in the cryogenic regime. When considered together with the temperature dependence of resistivity and Hall carrier concentration, these results provide a consistent picture of charge transport in HPGe across the cryogenic to near-room-temperature range.

\section{Discussion}

The combined analysis of resistivity, Hall carrier concentration, and Hall mobility provides a unified understanding of charge transport in USD-grown high-purity germanium over the temperature range 2--300~K. The observed behavior is governed by the interplay between carrier statistics and multiple scattering mechanisms, whose relative importance evolves continuously with temperature.

At low temperatures ($T \lesssim 15$~K), transport is primarily limited by carrier freeze-out. The strong suppression of the effective Hall carrier concentration indicates incomplete dopant ionization, leading to carrier-density-limited conduction. In this regime, the resistivity is controlled predominantly by carrier statistics rather than mobility. Although the measured Hall mobility increases with decreasing temperature, it should be interpreted as an apparent transport parameter, as carrier freeze-out, impurity-band conduction, and possible hopping processes modify the relationship between the Hall response and true drift transport.

As temperature increases into the intermediate regime ($\sim 15$--$80$~K), the progressive ionization of acceptors increases the free carrier concentration while scattering processes begin to dominate transport. The separation of mobility curves among samples reflects variations in effective impurity concentration and compensation, indicating that impurity-related scattering remains significant. This regime represents a crossover where both carrier density and scattering mechanisms contribute comparably, leading to the non-monotonic behavior observed in resistivity.

At temperatures ($\sim 80$--$200$~K), transport becomes mobility-limited, and the observed approximate scaling $\mu \propto T^{-3/2}$ is broadly consistent with transport behavior expected when acoustic-phonon-related scattering becomes increasingly important. The convergence of mobility across samples further supports the reduced influence of impurity scattering at elevated temperatures.

In the high-temperature regime ($T \gtrsim 200$~K), the decrease in resistivity reflects the onset of mixed transport, where thermally generated intrinsic carriers begin to contribute to conductivity. Under these conditions, the Hall coefficient increasingly reflects combined electron and hole transport rather than a pure single-carrier response. Consequently, the extracted effective Hall carrier concentration should be interpreted qualitatively in this temperature range rather than as a true single-carrier hole concentration. Because the intrinsic carrier concentration remains comparable to or smaller than the effective Hall carrier concentration in the present samples, the system does not reach a purely intrinsic regime within the measured temperature range. Instead, the observed behavior reflects mixed transport involving both impurity-derived and thermally generated carriers.

The temperature dependence of mobility is well captured by Matthiessen’s-rule-inspired phenomenological mobility model. While this approach provides a compact description of the dominant trends, the extracted parameters should be interpreted as phenomenological quantities rather than unique representations of individual scattering mechanisms. In particular, the limited sensitivity to separate impurity-related contributions, together with the limited number of experimental temperature points available for each sample, highlights the inherent difficulty of isolating microscopic processes in high-purity materials. Consequently, the model should not be regarded as a quantitative scattering analysis, but rather as a compact phenomenological framework for describing the overall mobility evolution and comparing transport behavior among samples.

Systematic variations among samples further emphasize the role of residual impurities and compensation. Samples with lower intermediate-temperature effective Hall carrier concentration generally exhibit higher mobility, suggesting a reduced overall influence of impurity-related scattering and compensation effects. These differences demonstrate the strong sensitivity of transport properties to small variations in impurity content, even within nominally high-purity material.
In addition to possible bulk impurity and compensation variations, geometric and surface-related effects may also contribute to the observed differences between sample groups. The A-series samples are substantially thinner ($\sim 118~\mu$m) than the B-series samples ($\sim 520$--$620~\mu$m), which may enhance the relative influence of surface conduction, contact geometry, polishing damage, surface depletion, or etching-related effects on the measured transport properties. Consequently, some portion of the observed sample-to-sample variation may reflect combined bulk and surface-sensitive transport contributions rather than purely intrinsic bulk impurity differences. Nevertheless, the overall systematic trends in resistivity, Hall carrier concentration, and mobility remain broadly consistent with variations in effective impurity concentration and compensation. Accordingly, the present measurements do not permit a complete separation of bulk impurity effects from surface-sensitive transport contributions. This limitation is expected to be most significant for the thinner A-series samples, where surface depletion, polishing damage, etching-related modifications, and contact geometry may have a proportionally larger influence on the measured transport properties.

From a detector perspective, these results have direct implications for cryogenic HPGe operation. Carrier freeze-out at low temperatures modifies the ionized impurity concentration, thereby affecting space-charge distributions, depletion characteristics, and electric-field formation. The resulting impact on detector bias requirements depends on detector geometry, net impurity concentration, contact design, and operating temperature. At the same time, increased mobility enhances charge collection speed, but the reduced carrier density and modified transport mechanisms can influence charge collection efficiency and noise performance. Residual impurities and compensation further affect charge drift uniformity and electronic noise, which are critical for low-threshold and low-noise detector applications.

\subsection{Comparison with Previous Germanium Transport Studies}
\label{subsec:comparison_literature}
\begin{table*}[!t]
\centering
\small
\caption{Comparison of representative transport parameters reported for germanium-based materials and the present USD-grown HPGe crystals. Reported ranges correspond to the temperature intervals investigated in the respective studies and are intended for qualitative comparison.}
\label{tab:comparison}

\begin{tabular}{cccccc}
\hline\hline
Reference &
Material Type &
$T$ (K) &
$\mu_H$ (cm$^2$ V$^{-1}$ s$^{-1}$) &
Effective Hall carrier concentration (cm$^{-3}$) &
$\rho$ ($\Omega\cdot$cm) \\
\hline

Koenig et al.~\cite{koenig1962electrical} &
n-type Ge &
4--25 &
$\sim1\times10^{5}$--$8.5\times10^{5}$ &
$\sim2\times10^{13}$ &
$\sim0.6$--$5$ \\

Palleti et al.~\cite{palleti2024properties} &
Bulk Compensated HPGe &
77 &
$\sim4.6\times10^{4}$ &
$\sim4.5\times10^{10}$ &
$\sim4.4\times10^{3}$ \\

\textbf{Present Work} &
\textbf{Bulk HPGe} &
\textbf{2--300} &
\textbf{$\sim1\times10^{3}$--$4.5\times10^{6}$} &
\textbf{$\sim1\times10^{10}$--$1.3\times10^{14}$} &
\textbf{$\sim5$--$10^{3}$} \\

\hline\hline
\end{tabular}

\end{table*}

To place the present results in context, Table~\ref{tab:comparison} compares representative transport properties of the USD-grown HPGe crystals with values reported in previous germanium transport studies. The comparison provides a useful benchmark for the mobility, carrier concentration, and resistivity ranges achieved in the present material.

The room-temperature Hall mobility of the USD-grown crystals is of order $10^{3}$ cm$^{2}$V$^{-1}$s$^{-1}$, consistent with values reported for high-purity germanium. At cryogenic temperatures, the Hall mobility reaches values as high as $\sim4.5\times10^{6}$ cm$^{2}$V$^{-1}$s$^{-1}$, comparable to the highest values reported in the literature. The effective Hall carrier concentration spans approximately $10^{10}$--$10^{14}$ cm$^{-3}$ across the investigated temperature range, reflecting the strong carrier freeze-out behavior observed at low temperatures. Likewise, the measured resistivity range extends from a few $\Omega\cdot$cm to approximately $10^{3}$ $\Omega\cdot$cm, consistent with highly compensated high-resistivity germanium.
These comparisons show that the measured cryogenic mobilities are within the range reported for high-quality germanium materials and support the relevance of USD-grown crystals for future detector-development studies.

Overall, the results establish a consistent physical picture of charge transport in HPGe, characterized by a continuous transition from carrier-density-limited transport at low temperatures to phonon-limited transport at high temperatures, with a mixed intermediate regime governed by both carrier ionization and scattering. This picture provides a useful basis for interpreting the overall transport behavior in HPGe crystals and for optimizing detector performance across a wide range of operating conditions.

\section{Conclusion}

We have presented a systematic experimental study of temperature-dependent charge transport in USD-grown p-type high-purity germanium over the range 2--300~K using combined Hall-effect and four-probe resistivity measurements. The joint analysis of resistivity, Hall carrier concentration, and Hall mobility provides a coherent and self-consistent picture of transport across cryogenic to near-room-temperature conditions.

The observed behavior is governed by the interplay between carrier ionization and scattering mechanisms. At low temperatures, carrier freeze-out leads to carrier-density-limited transport, while at intermediate temperatures both carrier ionization and impurity-related scattering contribute. At higher temperatures, transport becomes predominantly phonon-limited, and at the highest temperatures investigated, intrinsic carriers begin to contribute, resulting in a mixed transport regime rather than a purely intrinsic one.

The different observables exhibit distinct sensitivities to the underlying physics: resistivity reflects the combined influence of carrier density and mobility, Hall carrier concentration captures the freeze-out and extrinsic regimes, and Hall mobility reveals the transition from impurity-influenced to phonon-limited transport. Together, these measurements establish a unified framework for understanding charge transport in high-resistivity USD-grown germanium crystals.

Systematic variations among samples demonstrate the strong sensitivity of transport properties to residual impurities and compensation. The Matthiessen’s-rule-inspired phenomenological mobility model provides a compact description of the measured mobility trends, although the extracted parameters should be interpreted as effective fitting parameters rather than uniquely microscopic quantities.

These results are relevant for future HPGe detector-material development and transport modeling. Carrier freeze-out at cryogenic temperatures influences the ionized impurity distribution, depletion characteristics, and electric-field formation, while mobility and impurity scattering affect charge-collection dynamics and electronic noise. The measured transport properties therefore provide experimental guidance for future optimization of high-resistivity, high-purity germanium materials and for the development of detector-grade HPGe crystals intended for low-threshold and low-noise applications.

Overall, this work establishes a consistent experimental and phenomenological framework for charge transport in high-purity germanium, providing both fundamental insight and practical guidance for future optimization of germanium materials and cryogenic detector technologies.

\section*{Author Contributions}
N.~Budhathoki performed sample fabrication, surface processing, data analysis, and manuscript preparation. D.-M.~Mei provided conceptual guidance, supervision, and theoretical interpretation. A.~Rajbanshi contributed to data measurement, data analysis, and manuscript preparation. R. ~Jin provided supervision and theoretical interpretation.

\section*{Acknowledgments}
This work was supported in part by the National Science Foundation (NSF)  under Grants No.~NSF OISE-1743790, NSF PHYS 2117774, NSF OIA 2427805, NSF PHYS-2310027, and NSF OIA-2437416, and by the U.S. Department of Energy (DOE) under Grants No.~DE-SC0024519 and DE-SC0004768. Additional support was provided by a research center funded by the State of South Dakota. Work by A.R. and R.J. was supported by Grant No. DE-SC0024501 funded by the U.S. Department of Energy, Office of Science.

\section*{Data Availability}
The data that support the findings of this study are available from the 
corresponding author upon reasonable request.
\bibliographystyle{apsrev4-2}
\bibliography{references}
\end{document}